\documentstyle [12pt,twoside]{article}
\newskip\humongous \humongous=0pt plus 1000pt minus 1000pt
\def\caja{\mathsurround=0pt}

\newif\ifdtup
\def\panorama{\global\dtuptrue \openup2\jot \caja
        \everycr{\noalign{\ifdtup \global\dtupfalse
        \vskip-\lineskiplimit \vskip\normallineskiplimit
        \else \penalty\interdisplaylinepenalty \fi}}}

\def\eqalignnotwo#1{\panorama \tabskip=\humongous
        \halign to\displaywidth{\hfil$\displaystyle{##}$
        \tabskip=0pt&$\displaystyle{{}##}$
        \tabskip=\humongous&\llap{$##$}\tabskip=0pt
        \tabskip=0pt&$\displaystyle{{}##}$\hfil
        \crcr#1\crcr}}

\def\begintitle#1#2#3#4
        {\begin{titlepage}
         \centerline{#1 \hfill UMDGR-93-#2}
         \begin{center}\vglue .4in
         {\large\bf #3}\\[.4in]
         {\bf #4}\\[.2in]
         {\it Department of Physics}\\
         {\it University of Maryland, College Park, MD 20742}\\[.4in]
         {\bf ABSTRACT}\\
         \end{center}
         \begin{quotation}}
\def\endtitle
         {\end{quotation}
          \end{titlepage}
          \newpage}

\def\cC{{\cal C}}\def\cD{{\cal D}}
\def\cG{{\cal G}}
\def\cL{{\cal L}}

\def\cS{{\cal S}}
\def\cV{{\cal V}}\def\cW{{\cal W}}

\def\d{\delta}\def\e{\epsilon}

\def\s{\sigma}

\def\L{\Lambda}\def\S{\Sigma}

\def\pd{\partial}

\def\utw#1{\rlap{\lower1ex\hbox{$\sim$}}#1{}}
\def\pb#1{\rlap{\lower1ex\hbox{$\leftarrow$}}#1{}}
\def\pf#1{\rlap{\lower1ex\hbox{$\rightarrow$}}#1{}}

\def\Tr{{\rm Tr}}
\def\3#1{{}^3\!#1}\def\4#1{{}^4\!#1}\def\+#1{{}^+\!#1}\def\-#1{{}^-\!#1}
\def\*#1{{}^*\!#1}

\begin{document}

\begintitle{June 1993}{204}{\hfil ON THE CONSTRAINT ALGEBRA\hfil\break
OF DEGENERATE RELATIVITY}{Joseph
D. Romano\footnote{Address after September 1, 1993: Physics Department, The
University of Utah, 201 James Fletcher Building, Salt Lake City, Utah 84112}}
As shown by Ashtekar in the mid 80's, general relativity can be extended to
incorporate degenerate metrics. This extension is not unique, however, as one
can change the form of the hamiltonian constraints and obtain an {\it
alternative} degenerate extension of general relativity that disagrees with
Ashtekar's original theory when the triads vectors are degenerate.  In this
paper, the constraint algebra of a particular alternative theory is explicitly
evaluated and compared with that of Ashtekar's original degenerate extension. A
generic classification of the difference between the two theories is given in
terms of the degeneracy and surface-forming properties of the triad vectors.
(This classification is valid when the degeneracy and surface-forming
properties of the triad vectors is the same everywhere in an open set about a
point in space.) If the triad vectors are degenerate and surface-forming, then
all the secondary constraints of the alternative degenerate extension are
satisfied as a consequence of the primary constraints, and the constraints of
this theory are weaker than those of Ashtekar's. If the degenerate triad
vectors are not surface-forming, then the first secondary constraint of the
alternative theory already implies equivalence with Ashtekar's degenerate
extension. What happens when the degeneracy and surface-forming properties of
the triad vectors change from point to point is an open question.
\vskip .5cm

PACS: 04.20, 03.50
\endtitle

\noindent{\bf 1. Introduction}
\vskip .5cm

General relativity can be extended to incorporate degenerate metrics.  Indeed,
as shown by Ashtekar \cite{Ashtekar1} in the mid 80's, the hamiltonian
constraints of general relativity simplify when expressed in terms of a complex
$SO(3)$ connection and a density-weighted spatial triad. The constraints are
{\it polynomial} in these canonical variables, and the theory is well-defined
even if the spatial metric is degenerate.  Witten \cite{Witten} used this
result to quantize 2+1 gravity, and Horowitz \cite{Horowitz} has shown that by
extending general relativity to include degenerate metrics, topology change is
allowed classically. In fact, much of the renewed interest in the
non-perturbative canonical quantization program for 3+1 gravity is due to
Ashtekar's degenerate extension of general relativity \cite{Ashtekar2}.

More recently, Jacobson and Romano \cite{J&R} have shown that there exists a
natural {\it alternative} choice for the form of the hamiltonian constraints
that leads to a theory that agrees with general relativity (GR) for
non-degenerate metrics, but differs in the degenerate sector from Ashtekar's
original degenerate extension. The Poisson bracket (PB) algebra of the
alternative constraints fails to close in the degenerate sector due to
structure functions that involve the inverse of the spatial triad. The
alternative theory has to be supplemented with an apparently infinite number of
secondary constraints, which together with the primary constraints are shown to
be first class.  All of these constraints are implied by, but do not imply,
Ashtekar's original form of the constraints, and so give rise to a different
degenerate extension of GR.

One of the most puzzling and unfamiliar features of the alternative degenerate
extension is its apparently {\it infinite} number of secondary
constraints. It would seem that such a theory, with an infinite number of
constraints per space point, would have no local degrees of freedom. Yet, as
mentioned above and as shown in \cite{J&R}, the phase space of the alternative
theory is actually {\it larger} than the phase space of Ashtekar's theory.  One
can write down solutions to all the secondary constraints that fail to be
solutions to Ashtekar's original constraints.  This suggests that the infinite
number of constraints of the alternative theory are not all independent, but
without an explicit expression for the higher-order secondary constraints, one
does not know how to do the counting.

The purpose of this paper is to resolve this puzzle. We will find explicit
expressions for all the higher-order secondary constraints.\footnote{Since we
will be interested only in obtaining a closed PB algebra for the alternative
degenerate extension, we will evaluate the PBs {\it modulo} previous
constraints. We will not find explicit expressions for the structure functions
of this theory.} These expressions can be written in terms of nested
commutators of the triad vectors contracted with the vector constraint of
Ashtekar's original degenerate extension.  This new result will enable us to
show that the ``infinite" tower of secondary constraints for the alternative
theory is not infinite, but actually collapses to a finite number of
constraints. We will be able to generically\footnote{This classification is
valid when the degeneracy and surface-forming properties of the triad vectors
is the same everywhere in an open set about a point in space.} classify the
difference between Ashtekar's original degenerate extension of GR and the
alternative degenerate extension in terms of the degeneracy and surface-forming
properties of the triad vectors. In particular, we shall see that if the triad
vectors are degenerate and surface-forming, then all the secondary constraints
of the alternative degenerate extension are satisfied as a consequence of the
primary constraints, and the constraints of this theory are weaker than those
of Ashtekar's. If the degenerate triad vectors are not surface-forming, then
the first secondary constraint of the alternative theory already implies
equivalence with Ashtekar's degenerate extension. What happens when the
degeneracy and surface-forming properties of the triad vectors change from
point to point is an open question.

The rest of this paper is organized as follows: In section 2, I will briefly
review the hamiltonian formulation of Ashtekar's original degenerate extension
of GR \cite{Ashtekar1}.  My intent in this section is to give just enough
detail to write down the PB algebra of the constraints. Readers interested in a
more thorough discussion of Ashtekar's theory, and Ashtekar variables in
general, should see \cite{Ashtekar2} and references mentioned therein. In
section 3, I will describe the alternative degenerate extension of GR
originally discussed in \cite{J&R}. I will reproduce many of results found in
\cite{J&R}, but using a different notation (triads versus spinors) and from a
slightly different point of view.  For instance, the method I use to calculate
the PBs for the alternative theory yields an explicit expression for all the
higher-order secondary constraints---expressions that were lacking in
\cite{J&R}. In section 4, I will give a generic classification of the
difference between the two degenerate extensions of GR and show how the
apparent infinite tower of secondary constraints for the alternative theory
collapses to a finite number of constraints. As mentioned above, this
classification is valid when the degeneracy and surface-forming properties of
the triad vectors is the same everywhere in an open set about a point in space.
In section 5, I briefly summarize the main results and conclude by raising some
questions regarding those cases where the degeneracy and surface-forming
properties of the triad vectors change from point to point. An appendix gives
details of the calculation of the PBs for the alternative degenerate extension
discussed in section 3.

\vskip .5cm
\noindent{\bf 2. Ashtekar's original degenerate extension of GR}
\vskip .5cm

In Ashtekar's original degenerate extension of GR \cite{Ashtekar1}, the phase
space variables consist of a complex $SO(3)$ connection $A_a^i$ and a
density-weighted (+1) spatial triad $E_i^a$. Both are defined on a 3-manifold
$\S$. (In what follows, $a,b,c,\cdots $ denote spatial indices, while
$i,j,k,\cdots$ denote internal $SO(3)$ indices.) These variables are subject
to the constraints
$$\eqalignnotwo{\cG_i&\equiv\cD_a E_i^a=0&(1a)\cr
\cS&\equiv\e^{ij}{}_k E_i^a E_j^b F_{ab}^k=0&(1b)\cr
\cV_a&\equiv E_j^b F_{ab}^j=0&(1c)\cr}$$
where $\cD_a$ is the gauge-covariant derivative operator associated with
$A_a^i$, and $F_{ab}^i$ is its associated curvature.  Explicitly, $\cD_a
E_i^a=\pd_a E_i^a + \e_{ij}{}^k A_a^j E_k^a$ and $F_{ab}^i=2\pd_{[a}A_{b]}^i +
\e^i{}_{jk} A_a^j A_b^k$, where $\pd_a$ is some fixed, flat derivative operator
that ignores internal indices. $\e_{ijk}$ is the Levi-Civita symbol for the
internal space.  The constraints $(1a,b,c)$ are called the Gauss, scalar, and
vector constraints, respectively.  Since they are polynomial in $A_a^i$ and
$E_i^a$,  this theory extends GR to incorporate degenerate spatial metrics
$q q^{ab}\equiv E^a_i E^{bi}$.

To evaluate the PB algebra of the constraints, I will first define constraint
functionals $\cG_\L$, $\cS_N$, and $\cV_{\vec N}$ by smearing $\cG_i$, $\cS$,
and $\cV_a$ with appropriate test fields on $\S$.  By doing this, we obtain
real-valued (rather than tensor-valued) functions on phase space. Explicitly,
$$\eqalignnotwo{\cG_\L&\equiv\int_\S \L^i\cD_a E_i^a=0&(2a)\cr
\cS_N&\equiv\int_\S N\e^{ij}{}_k E_i^a E_j^b F_{ab}^k=0&(2b)\cr
\cV_{\vec N}&\equiv\int_\S N^a E_j^b F_{ab}^j=0&(2c)\cr}$$
where, for instance, $N$ is a scalar density of weight $-1$.  With this choice
for $N$, the integrand of Eq.~($2b$) is a scalar density of $+1$, which can be
integrated over $\S$ without the need of any additional structure. Since we
will concentrate only on the constraint algebra and not on dynamics, we can
assume for asymptotically flat $\S$ that the smearing fields fall-off
sufficiently fast so that the above constraint functionals are differentiable
and boundary terms can be ignored. For spatially compact $\S$, these issues do
not arise since $\pd\S=0$.

As shown in \cite{Ashtekar1}, the PB algebra of the constraints is {\it
closed}. That is, the PB of any two constraint functionals, Eqs.~($2a,b,c$),
vanish modulo themselves.  Explicitly, if one takes $\{E_i^a(x),A_b^j(y)\}
=\d_i^j\d_b^a\d^3(x,y)$ as the fundamental PB, one can show that
$$\eqalignnotwo{\{\cG_\L,\cG_{\L'}\}&=-\cG_{[\L,\L']}\approx 0&(3a)\cr
\{\cG_\L,\cV_{\vec M}\}&=0&(3b)\cr
\{\cG_\L,\cS_M\}&=0&(3c)\cr
\{\cV_{\vec N},\cV_{\vec M}\}&=-\cV_{[\vec N,\vec M]}+\cG_{N^a M^b F_{ab}}
\approx 0&(3d)\cr
\{\cS_N,\cV_{\vec M}\}&=\cS_{\cL_{\vec M} N}+\cG_{2N M^a[E^b,F_{ab}]}\approx 0
&(3e)\cr
\{\cS_N,\cS_M\}&=\cV_{4\vec E{}_i E^{bi}(N\pd_b M-M\pd_b N)}\approx 0&(3f)
\cr}$$
where $[\L,\L']\equiv\e^i{}_{jk}\L^j\L^k$, $[\vec N,\vec M]^a\equiv N^b\pd_b
M^a-M^b\pd_b N^a$, and $[E^b,F_{ab}]^i\equiv\e^{ij}{}_k E_j^b F_{ab}^k$.
Since the RHSs of all these equations are weakly zero, there are {\it no}
secondary constraints.

\vskip .5cm
\noindent{\bf 3. Alternative degenerate extension of GR}
\vskip .5cm

The alternative degenerate extension of GR originally discussed in \cite{J&R}
is defined by the same phase space variables $(A_a^i,E_i^a)$ as in Ashtekar's
theory, but they are subject to a {\it different} set of constraints. The Gauss
and scalar constraints Eqs.~($1a,b$) are the same, but the vector constraint
$\cV_a=0$ is replaced by
$$\cW_i\equiv E_i^a \ \cV_a= E_i^a E_j^b F_{ab}^j=0.\eqno(4)$$
$\cW_i=0$ will be called the {\it weak} vector constraint since it is implied
by, but does not in general imply, the original vector constraint $\cV_a=0$.
Since the weak vector constraint is polynomial in $A_a^i$ and $E_i^a$, the
alternative theory defined by the primary constraints Eqs.~($1a,b$) and (4)
also extends GR to incorporate degenerate spatial metrics.  The constraints of
the two theories are equivalent if and only if the triad vectors $E_i^a$ are
non-degenerate.\footnote{In $SL(2,C)$ spinor notation (see \cite{Ashtekar1} or
\cite{J&R}), the phase variables are denoted by $A_a{}^{AB}$ and $\s^a{}_{AB}$,
and the scalar and vector constraints ($1b,c$) have the form
$\cS\equiv\Tr(\s^a\s^b F_{ab})=0$ and $\cV_a\equiv\Tr(\s^b F_{ab})=0$. These
can be written as a {\it combined} constraint $\cC_{AB}\equiv(\s^a\s^b
F_{ab})_{AB}=0$ with only internal indices, which in terms of $\cV_a$ and $\cS$
equals
$$\cC_{AB}={1\over 2}\e_{AB} \ \cS + \s^a{}_{AB} \ \cV_a.$$
Thus, $\cC_{AB}=0$ if and only if $\cS=0$ and $\cW_{AB}\equiv\s^a{}_{AB} \
\cV_a=0$. The constraint $\cW_{AB}=0$ corresponds to the constraint $\cW_i=0$
in the triad notation of this paper. One had originally hoped that by writing
the four diffeomorphism constraints in this combined form, the PB algebra of
the constraints would simplify.  Such a simplification occurs in 2+1 gravity;
the PB algebra there being $ISO(2,1)$. Unfortunately, such a simplification
does not occur for the 3+1 theory, even if one restricts attention to
non-degenerate triad vectors \cite{J&R}.}

Unlike the PB algebra of Ashtekar's original theory, the PB algebra of the
alternative degenerate extension is {\it not} closed \cite{J&R}. The PBs of the
Gauss and scalar constraint functionals weakly vanish modulo the Gauss and weak
vector constraints (see Eqs.~($3a,c,f$)), but the PBs involving the weak vector
constraint are not all weakly zero. Defining
$$\cW_N\equiv\int_\S N^i E_i^a E_j^b F_{ab}^j=0,\eqno(5)$$
where $N^i$ is a smearing field of density weight $-1$, one finds
$$\eqalignnotwo{\{\cW_N,\cW_M\}&\approx\cV_{[\vec N(E),\vec M(E)]}&(6a)\cr
\{\cS_N,\cW_M\}&\approx\cV_{N M^i\e_i{}^{jk}[\vec E_j,\vec E_k]}&(6b)\cr
\{\cG_\L,\cW_M\}&\approx 0&(6c)\cr}$$
where $\vec N(E)$ and $\vec M(E)$ denote the field-dependent vector fields
$N^a(E)\equiv N^i E_i^a$ and $M^a(E)\equiv M^i E_i^a$, and $[E_i,E_j]^a\equiv
E_i^b\pd_b E_j^a-E_j^b\pd_b E_i^a$ is the {\it commutator}\footnote{The gauge
dependence of this so-called commutator of vector densities will be discussed
below, in the second-to-last paragraph of this section.} of $E_i^a$ and
$E_j^a$. Equations (6) hold modulo the Gauss, scalar, and weak vector
constraints. (See the appendix of this paper or \cite{J&R} for a detailed
calculation of the above PBs.) The RHS of Eqs.~(6) are weakly zero for all
smearing fields if and only if
$$\cW_{ij}\equiv[E_i,E_j]^a\cV_a\approx 0.\eqno(7)$$
This is a {\it secondary} constraint.

Since the primary constraints are not closed under PB, we must repeat the above
process.  That is, we must also require that the PBs of $\cW_N$, $\cS_N$, and
$\cG_\L$ with the secondary constraint Eq.~(7) be weakly equal to zero if we
eventually hope to obtain a consistent theory with a closed PB algebra. Since
the constraint functional
$$\int_\S N^{ij}\cW_{ij}\equiv\int_\S N^{ij}[E_i,E_j]^a\cV_a\eqno(8)$$
can be written as a sum of terms, each of the form
$$\int_\S M^{[i} L^{j]}\cW_{ij}=\int_\S M^i L^j[E_i,E_j]^a\cV_a\approx
\cV_{[\vec M(E),\vec L(E)]},\eqno(9)$$
it suffices to calculate the PBs of $\cW_N$, $\cS_N$, and $\cG_\L$ with
$\cV_{[\vec M(E),\vec L(E)]}$. Evaluating these higher-order PBs leads to what
I will call a {\it higher-order} secondary constraint.  Explicitly, one finds
$$\eqalignnotwo{\{\cW_N,\cV_{[\vec M(E),\vec L(E)]}\}&\approx\cV_{[\vec N(E),
[\vec M(E),\vec L(E)]]}&(10a)\cr
\{\cS_N,\cV_{[\vec M(E),\vec L(E)]}\}&\approx\cV_{N(M^iL^l-L^iM^l)\e_l{}^{jk}
[\vec E_i,[\vec E_j,\vec E_k]]}&(10b)\cr
\{\cG_\L,\cV_{[\vec M(E),\vec L(E)]}\}&\approx 0&(10c)\cr}$$
where $\approx 0$ now means $=0$ modulo $\cG_i=0$, $\cS=0$, $\cW_i=0$, and
$\cW_{ij}=0$. (Again, see the appendix for details involving the calculation of
the above PBs.) The RHS of the above equations are weakly zero for all smearing
fields if and only if
$$\cW_{ijk}\equiv[E_i,[E_j,E_k]]^a\cV_a\approx 0.\eqno(11)$$
This is a tertiary, or third-order, constraint. The explicit expression
Eq.~(11) for $\cW_{ijk}$ is a {\it new} result that was not known in
\cite{J&R}. By repeating this procedure, one finds $\cW_{ijkl}\approx 0, \cdots
,$ where $\cW_{ijkl}, \cdots ,$ are the higher-order generalizations of
Eqs.~(7) and (11).

Two remarks are in order. The first involves what I have been calling the
{\it commutator} of the triad vectors $E_i^a$ and $E_j^a$. As noted in the
sentence immediately following Eqs.~(6), $[E_i,E_j]^a$ is defined by
$$[E_i,E_j]^a\equiv E_i^b\pd_b E_j^a - E_j^b\pd_b E_i^a\eqno(12)$$
where $\pd_a$ is the fixed, flat derivative operator that we originally used to
define $\cD_a$ in terms of $A_a^i$. Since $E_i^a$ are vector {\it densities} of
weight $+1$, Eq.~(12) depends on the choice of $\pd_a$. In this sense, the
commutator of the triad vectors is gauge dependent. Nonetheless, the
constraints $\cW_{ij}\approx 0$, $\cW_{ijk}\approx 0, \cdots ,$ are independent
of the choice of $\pd_a$ {\it modulo previous constraints}. Moreover,
$[E_i,E_j]^a=C^k{}_{ij}E_k^a$ has the same geometrical interpretation as
surface-forming vector fields, again independent of the the choice of $\pd_a$.
This is because the commutator $[E_i,E_j]^a$ picks up terms proportional to
$E_i^a$ and $E_j^a$ under a change of derivative operator.

The second remark is that the totality of constraints for the alternative
degenerate extension of GR is {\it first class}. This was proved in \cite{J&R}
using the Jacobi identity
$$\{f,\{g,h\}\}+\{g,\{h,f\}\}+\{h,\{f,g\}\}=0\eqno(13)$$
for PBs. In fact, the proof given there showed that the first class nature of
the constraints does not depend on the explicit form of the higher-order
secondary constraints. As an illustration of this general result, consider the
PB of two secondary constraints $\cV_{[\vec N(E),\vec M(E)]}$ and $\cV_{[\vec
L(E),\vec K(E)]}$. Using Eqs.~(13), ($6a$), ($10a$), and their higher-order
generalizations, we find
$$\eqalignnotwo{\{\cV_{[\vec N(E),\vec M(E)]},\cV_{[\vec L(E),\vec K(E)]}\}
&\approx\{\{\cW_N,\cW_M\},\{\cW_L,\cW_K\}\}&\cr
&=\{\cW_K,\{\cW_L,\{\cW_N,\cW_M\}\}-\{\cW_L,\{\cW_K,\{\cW_N,\cW_M\}\}&\cr
&\approx\cV_{[\vec K(E),[\vec L(E),[\vec N(E),\vec M(E)]]]}- L\leftrightarrow
K&(14)\cr}$$
modulo first, second, and third-order constraints. Thus, the PB of two
secondary constraints is weakly equal to a sum of fourth-order constraints.

\vskip .5cm
\noindent{\bf 4. Generic classification: Redundancy of the higher-order
constraints}
\vskip .5cm

We are now ready to address the question regarding the redundancy of the
higher-order secondary constraints for the alternative degenerate extension of
GR.  In the process of answering it, we will obtain a generic classification of
the difference between the alternative theory and Ashtekar's original
degenerate extension of GR. As suggested by Eqs.~(7), (11), and their
higher-order generalizations, the relationship between the two theories depends
on the degeneracy and surface-forming properties of the triad vectors.

Let $x\in\S$ be any point of space, and assume that the degeneracy and
surface-forming properties of the triad vectors is the same in an open set
about $x$.  Then we have the following generic classification:
\begin{enumerate}
\item If the  triad vectors are non-degenerate, then $\cW_i=0$ is equivalent to
$\cV_a=0$, and $\cW_{ij}=0, \cdots ,$ are automatically satisfied.  There is
nothing surprising here.  We already knew that both theories agree in the
non-degenerate sector.
\item If the triad vectors are degenerate and surface-forming, then although
$\cW_i=0$ does not imply $\cV_a=0$, the secondary constraints $\cW_{ij}=0,
\cdots ,$ are automatically satisfied as a consequence of $\cW_i=0$. This is
because $[E_i,E_j]^a, \cdots ,$ are proportional to $E_k^a$, and $E_k^a \
\cV_a=\cW_k =0$. Thus, the alternative degenerate extension of GR is different
from Ashtekar's original degenerate extension.
\item If the triad vectors are degenerate but are {\it not} surface-forming,
then $\cW_i=0$ and the first secondary constraint $\cW_{ij}=0$ already imply
$\cV_a=0$. (The higher-order secondary constraints $\cW_{ijk}=0, \cdots ,$
are automatically satisfied as well.) This is because $E_i^a$ and $[E_i,E_j]^a$
are linearly independent, so $\cW_i\equiv E_i^a \cV_a=0$ and $\cW_{ij}\equiv
[E_i,E_j]^a\cV_a=0$ imply $\cV_a=0$. Thus, for this case, the alternative
degenerate extension of GR is equivalent to Ashtekar's theory.
\end{enumerate}

As explicit example of case 2 above, one can consider
$$E_i^a\equiv\left((\pd/\pd x)^a,(\pd/\pd y)^a,0\right)\quad{\rm and}\quad
A_a^i\equiv\left(x(dz)_a,y(dz)_a,0\right)\eqno(15)$$
where $(x,y,z)$ are coordinates about some point in space. For such initial
data, $\cG_i=0$, $\cS=0$, $\cW_i=0$, $\cW_{ij}=0, \cdots ,$ but
$\cV_a=-2(dz)_a\not=0$. Thus, these phase space variables are a solution to
the alternative theory, but fail to be a solution of Ashtekar's
theory.\footnote{Under the infinitesimal evolution generated by the scalar
constraint functional $\cS_N$, one finds $\dot A_a^i=0$ (for $i=1,2,3$), $\dot
E_1^a = 0$, $\dot E_2^a=0$, and $\dot E_3^a= 2(\pd N/\pd y)(\pd/\pd x)^a-2(\pd
N/\pd x) (\pd/\pd y)^a$, where $\dot f\equiv\{\cS_N,f\}$. Thus, for $N$
independent of $x$ and $y$ (in particular, for $N=1$), this degenerate
solution, Eq.~(15), of the initial value constraints of the alternative theory
is constant in time.}

\vskip .5cm
\noindent{\bf 5. Conclusion}
\vskip .5cm

By finding an explicit expression for the higher-order secondary constraints of
the alternative degenerate extension of GR, we have been able to resolve the
puzzle of the ``infinite" tower of secondary constraints originally raised in
\cite{J&R}.  From Eqs.~(7), (11), etc.~one sees that the secondary constraints
and their higher-order generalizations are nothing more than the vector
constraint of Ashtekar's original degenerate extension contracted with nested
commutators of the triad vectors. Whenever the triad vectors and any of the
nested commutators span the tangent space at a point in space, then $\cW_i=0$,
$\cW_{ij}=0, \cdots ,$ imply $\cV_a=0$ and the two theories are equivalent.
More generally, the classification given in section 4 completely characterizes
the difference between the two degenerate extensions of GR when  the degeneracy
and surface-forming properties of triad vectors is the same everywhere in an
open set about a point in space.

But what about those cases where the degeneracy and surface-forming properties
of the triad vectors change from point to point?  What can one say about the
relationship between the two degenerate extensions in these cases?
Unfortunately, I do not know the general answer to these questions at present.
Only when the triad vectors are continuous and degenerate on a set of measure
zero, can I make a definite statement.  For such a triad, the two theories are
equivalent. This is because $\cW_i\equiv E_i^a\cV_a=0$ implies $\cV_a=0$
wherever $E_i^a$ is non-degenerate, while continuity of $\cV_a$ implies
$\cV_a=0$ elsewhere.  The classification for the opposite case (where the triad
vectors have {\it support} only on a set of measure zero) is yet to be
understood.

\vskip .5cm
\noindent ACKNOWLEDGEMENTS
\vskip .3cm

I would like to thank Ted Jacobson for the numerous discussions I had with him
throughout the course of this work.  I would also like to thank Riccardo
Capovilla, Jonathan Simon, and Ranjeet Tate for their helpful remarks,
especially related to the calculation of the PBs for the alternative degenerate
extension. This work was supported by NSF grant PHY91-12240.

\vskip .5cm
\noindent{\bf Appendix: Calculation of the PBs for the alternative degenerate
extension}
\vskip .5cm

In this appendix, I will describe in detail a method of calculating the PBs of
$\cW_N$, $\cS_N$, and $\cG_\L$ with the weak vector constraint $\cW_M$ of the
alternative degenerate extension of GR. This method of calculation takes
advantage of the PBs (Eqs.~(3)) for Ashtekar's original degenerate extension.
This method of calculation can also be generalized to give the PBs of $\cW_N$,
$\cS_N$, and $\cG_\L$ with the secondary constraint functional $\cV_{[\vec
M(E),\vec L(E)]}$ and with all higher-order secondary constraints. (The
necessary steps are sketched in the final paragraph of the appendix.)  Since we
will be interested only in obtaining a closed PB algebra for the alternative
degenerate extension, we will evaluate the PBs {\it modulo} previous
constraints.  We will not find explicit expressions for the structure functions
of this theory.

The key to this method of calculation is the equality $\cW_M=\cV_{\vec M(E)}$.
Using this equality, we can write
$$\eqalignnotwo{{\d\cW_M\over\d E_i^a}&={\d\cV_{\vec M}\over\d E_i^a}
\Bigg|_{\vec M(E)} + \cV_c \ {\pd M^c(E)\over\pd E_i^a}&(A.1a)\cr
{\d\cW_M\over\d A_a^i}&={\d\cV_{\vec M}\over\d A_a^i}\Bigg|_{\vec M(E)}
&(A.1b)\cr}$$
where $(\d\cV_{\vec M}/\d E_i^a)|_{\vec M(E)}$ denotes the variation of
$\cV_{\vec M(E)}$ with respect to $E_i^a$ holding $A_a^i$ fixed and treating
$M^a(E)\equiv M^i E_i^a$ as if it were {\it independent} of $E_i^a$.  A
similar definition applies to $(\d\cV_{\vec M}/\d A_a^i)|_{\vec M(E)}$.
Thus,
$$\eqalignnotwo{\{\cW_N,\cW_M\}&\equiv\{\cV_{\vec N(E)},\cV_{\vec M(E)}\}&\cr
&=\int_\S{\d\cW_N\over\d E_i^a}{\d\cW_M
\over\d A_a^i}-{\d\cW_M\over\d E_i^a}{\d\cW_N\over\d A_a^i}&\cr
&=\int_\S\Bigg({\d\cV_{\vec N}\over\d E_i^a}
\Bigg|_{\vec N(E)} + \cV_c \ {\pd N^c(E)\over\pd E_i^a}\Bigg){\d\cV_{\vec M}
\over\d A_a^i}\Bigg|_{\vec M(E)}-N\leftrightarrow M&\cr
&=\Big\{\cV_{\vec N}\Big|_{\vec N(E)},\cV_{\vec M}\Big|_{\vec M(E)}\Big\}
+\int_\S\Bigg(\cV_c \ {\pd N^c(E)\over\pd E_i^a}{\d\cV_{\vec M}\over\d A_a^i}
\Bigg|_{\vec M(E)}-N\leftrightarrow M\Bigg)&(A.2a)\cr}$$
where the first term on the last line denotes the PB of $\cV_{\vec N(E)}$
and $\cV_{\vec M(E)}$ treating $\vec N(E)$ and $\vec M(E)$ as if they
were independent of $E_i^a$. Similarly,
$$\eqalignnotwo{\{\cS_N,\cW_M\}&=\Big\{\cS_N,\cV_{\vec M}\Big|_{\vec M(E)}
\Big\}-\int_\S\cV_c \ {\pd M^c(E)\over\pd E_i^a}{\d\cS_N\over\d A_a^i}
&(A.2b)\cr
\{\cG_\L,\cW_M\}&=\Big\{\cG_\L,\cV_{\vec M}\Big|_{\vec M(E)}
\Big\}-\int_\S\cV_c \ {\pd M^c(E)\over\pd E_i^a}{\d\cG_\L\over\d A_a^i}.
&(A.2c)\cr}$$

To evaluate the above PBs (Eqs.~($A.2a,b,c$)), it is useful to note that the
first two terms on the last line of Eq.~($A.2a$) weakly cancel. This
cancellation holds modulo the Gauss and weak vector constraints. To see this,
recall that (see Eq.~($3d$))
$$\eqalignnotwo{\Big\{\cV_{\vec N}\Big|_{\vec N(E)},\cV_{\vec M}
\Big|_{\vec M(E)}\Big\}&=-\cV_{[\vec N(E),\vec M(E)]}+\cG_{N^a(E)M^b(E)F_{ab}}
&\cr
&\approx -\cV_{[\vec N(E),\vec M(E)]}.&(A.3)\cr}$$
In addition,
$${\d\cV_{\vec M}\over\d A_a^i}\Bigg|_{\vec M(E)}=2\cD_b(M^{[a}(E)E_i^{b]}).
\eqno(A.4)$$
If we expand the covariant derivative in Eq.~($A.4$) using the Leibnitz rule
and contract the resulting expression with $\pd N^c(E)/ \pd E_i^a=\d_a^c \
N^i$, we find
$${\pd N^c(E)\over\pd E_i^a}{\d\cV_{\vec M}\over\d A_a^i}\Bigg|_{\vec M(E)}=
[N(E),M(E)]^c+(\hbox{\rm terms proportional to $\cG_i$ and $E_i^c$}).
\eqno(A.5)$$
Thus,
$$\int_\S\cV_c \ {\pd N^c(E)\over\pd E_i^a}{\d\cV_{\vec M}\over\d A_a^i}
\Bigg|_{\vec M(E)}\approx\cV_{[\vec N(E),\vec M(E)]}\eqno(A.6)$$
modulo the Gauss and weak vector constraints. Combining Eqs.~($A.3$) and
($A.6$), we obtain
$$\{\cW_N,\cW_M\}\approx-\int_\S\cV_c \ {\pd M^c(E)\over\pd E_i^a}{\d\cV_{\vec
N}\over\d A_a^i}\Bigg|_{\vec N(E)}.\eqno(A.7a)$$
Similarly,
$$\eqalignnotwo{\{\cS_N,\cW_M\}&\approx-\int_\S\cV_c \ {\pd M^c(E)\over\pd
E_i^a}{\d\cS_N\over\d A_a^i}&(A.7b)\cr
\{\cG_\L,\cW_M\}&=-\int_\S\cV_c \ {\pd M^c(E)\over\pd E_i^a}{\d\cG_\L\over
\d A_a^i}.&(A.7c)\cr}$$
These hold modulo the Gauss and scalar constraints. (See Eqs.~($3e$) and
($3b$).)

To complete the calculation of the PBs, we need to evaluate various functional
derivatives.  These include
$$\eqalignnotwo{&{\d\cV_{\vec N}\over\d A_a^i}\Bigg|_{\vec N(E)}=2\cD_b
(N^{[a}(E)E_i^{b]})&(A.8a)\cr
&{\d\cS_N\over\d A_a^i}\quad=2\e_i{}^{jk}\cD_b(N E_j^{[a} E_k^{b]})&(A.8b)\cr
&{\d\cG_\L\over\d A_a^i}\quad=-\L^j\e_{ij}{}^k E_k^a.&(A.8c)\cr}$$
Since $M^a\equiv M^i E_i^a$, we also have
$${\pd M^c(E)\over\pd E_i^a} \ T_i^a=M^i T_i^c\equiv M^c(T)\eqno(A.9)$$
for any $T_i^a$ of density-weight $+1$. If we take $T_i^a$ to be the above
functional derivatives, Eqs.~($A.8$), and expand the covariant derivative in
Eqs.~($A.8a$) and ($A.8b$), we find
$$\eqalignnotwo{&{\pd M^c(E)\over\pd E_i^a}{\d\cV_{\vec N}\over\d A_a^i}
\Bigg|_{\vec N(E)}=-[N(E),M(E)]^c+(\hbox{\rm terms proportional to $\cG_i$
and $E_i^c$})&(A.10a)\cr
&{\pd M^c(E)\over\pd E_i^a}{\d\cS_N\over\d A_a^i}\quad=-N M^i\e_i{}^{jk}
[E_j,E_k]^c+(\hbox{\rm terms proportional to $\cG_i$ and $E_i^c$})&(A.10b)\cr
&{\pd M^c(E)\over\pd E_i^a}{\d\cG_\L\over\d A_a^i}\quad=(\hbox{\rm terms
proportional to $E_i^c$}).&(A.10c)\cr}$$
Thus,
$$\eqalignnotwo{\{\cW_N,\cW_M\}&\approx\cV_{[\vec N(E),\vec M(E)]}&(A.11a)\cr
\{\cS_N,\cW_M\}&\approx\cV_{N M^i\e_i{}^{jk}[\vec E_j,\vec E_k]}&(A.11b)\cr
\{\cG_\L,\cW_M\}&\approx 0&(A.11c)\cr}$$
modulo the Gauss, scalar, and weak vector constraints. These are Eqs.~(6) in
the main text.\footnote{Since the Gauss constraint generates internal gauge
transformations, one could have immediately written down
$\{\cG_\L,\cW_M\}=-\cW_{[\L,M]}$, where $[\L,M]^i\equiv\e^i{}_{jk}\L^j M^k$.
Thus, $\{\cG_\L,\cW_M\}\approx 0$ modulo the weak vector constraint. In fact,
the PB of the Gauss constraint with {\it any} constraint $\cC$ is either
identically zero (if $\cC$ is gauge invariant) or weakly zero modulo $\cC=0$
(if $\cC$ has internal indices).}

To evaluate the PBs of $\cW_N$, $\cS_N$, and $\cG_\L$ with the secondary
constraint functional $\cV_{[\vec M(E),\vec L(E)]}$, we can basically proceed
as above.  We simply replace $M^a(E)$ with the commutator $[M(E),L(E)]^a$ and
calculate the PBs modulo $\cG_i=0$, $\cS=0$, $\cW_i=0$, and $\cW_{ij}\equiv
[E_i,E_j]^a\cV_a=0$. We are allowed to make this simple replacement since
Eqs.~($A.7$), when written in terms of $\cV_{\vec M(E)}$,
$$\eqalignnotwo{\{\cW_N,\cV_{\vec M(E)}\}&\approx-\int_\S\cV_c \ {\pd M^c(E)
\over\pd E_i^a}{\d\cV_{\vec N}\over\d A_a^i}\Bigg|_{\vec N(E)}&(A.12a)\cr
\{\cS_N,\cV_{\vec M(E)}\}&\approx-\int_\S\cV_c \ {\pd M^c(E)\over\pd E_i^a}
{\d\cS_N\over\d A_a^i}&(A.12b)\cr
\{\cG_\L,\cV_{\vec M(E)}\}&=-\int_\S\cV_c \ {\pd M^c(E)\over\pd E_i^a}
{\d\cG_\L\over\d A_a^i}&(A.12c)\cr}$$
hold for {\it any} field-dependent smearing field $\vec M(E)$. In particular,
Eqs.~($A.12$) are valid when $\vec M(E)$ is replaced by the commutator
$[\vec M(E),\vec L(E)]$.  In addition, Eq.~($A.9$) should be replaced by
$$\eqalignnotwo{{\pd[M(E),L(E)]^c\over\pd E_i^a} \ T_i^a&=\Big[{\pd M(E)\over
\pd E_i^a} \ T_i^a,L(E)\Big]^c+\Big[M(E),{\pd L(E)\over\pd E_i^a} \ T_i^a
\Big]^c&\cr
&\equiv[M(T),L(E)]^c+[M(E),L(T)]^c&(A.13)\cr}$$
which together with Eqs.~($A.10$) imply
$$\eqalignnotwo{&{\pd[M(E),L(E)]^c\over\pd E_i^a}{\d\cV_{\vec N}\over\d A_a^i}
\Bigg|_{\vec N(E)}=-[N(E),[M(E),L(E)]]^c&\cr
&\hbox{\hskip 1.5 true in}+(\hbox{\rm terms proportional to $\cG_i$, $E_i^c$,
and $[E_i,E_j]^c$})&(A.14a)\cr
&{\pd[M(E),L(E)]^c\over\pd E_i^a}{\d\cS_N\over\d A_a^i}\quad=-N(M^iL^l-L^iM^l)
\e_l{}^{jk}[E_i,[E_j,E_k]]^c&\cr
&\hbox{\hskip 1.5 true in}+(\hbox{\rm terms proportional to $\cG_i$, $E_i^c$,
and $[E_i,E_j]^c$})&(A.14b)\cr
&{\pd[M(E),L(E)]^c\over\pd E_i^a}{\d\cG_\L\over\d A_a^i}\quad=(\hbox{\rm terms
proportional to $E_i^c$ and $[E_i,E_j]^c$}).&(A.14c)\cr}$$
Contracting the above expressions with $\cV_c$ and integrating over $\S$, we
find
$$\eqalignnotwo{\{\cW_N,\cV_{[\vec M(E),\vec L(E)]}\}&\approx\cV_{[\vec N(E),
[\vec M(E),\vec L(E)]]}&(A.15a)\cr
\{\cS_N,\cV_{[\vec M(E),\vec L(E)]}\}&\approx \cV_{N(M^iL^l-L^iM^l)\e_l{}^{jk}
[\vec E_i,[\vec E_j,\vec E_k]]}&(A.15b)\cr
\{\cG_\L,\cV_{[\vec M(E),\vec L(E)]}\}&\approx 0&(A.15c)\cr}$$
modulo $\cG_i=0$, $\cS=0$, $\cW_i=0$, and $\cW_{ij}=0$. These are Eqs.~(10) in
the main text. The PBs of $\cW_N$, $\cS_N$, and $\cG_\L$ with the higher-order
secondary constraints $\cV_{[\vec M(E),[\vec L(E),\vec K(E)]]}, \cdots ,$ are
obtained in a similar manner.

\end{document}